\documentclass[aps,pre,showpacs,twocolumn,floats,amsmath,amssymb,superscriptaddress,groupedaddress,10pt]{revtex4-1}

\usepackage{caption}
\usepackage{makeidx}
\usepackage{epsfig}
\usepackage{subfig}
\usepackage{graphics}
\usepackage{colortbl}
\usepackage{colordvi}
\usepackage{verbatim}
\usepackage{amsmath}
\usepackage{hyperref}
\usepackage{enumerate}
\usepackage{wrapfig}
\usepackage{bmpsize}

\begin{document}

\title{A Recipe for Composite Materials: An Approach through Fiber Bundle Model}

\author{Subhadeep Roy}
\email{sroy@imsc.res.in}
\affiliation{Institute of Mathematical Sciences, Taramani, CIT Campus, Chennai 600113}
\author{Sanchari Goswami}
\email{sanchari.goswami@bose.res.in}
\affiliation{S. N. Bose National Centre for Basic Sciences, Block -JD, Sector - III, Salt Lake, Kolkata 700098}


\begin{abstract}
Strengthening of materials and preventing abrupt fracture are really challenging jobs in the 
field of engineering and material science. Such problems can be resolved by using composite materials.
In this work, we have studied 
the fracture process of a composite material in light of fiber bundle model with different elastic constants 
as well as  different random threshold breaking strength of fibers. The critical 
width of the threshold distribution ($\delta_c$), for which
abrupt failure occurs, is studied both analytically and numerically 
with increasing number of components $(k)$ in the composite and it is  
shown that $\delta_c$ is 
inversely related to $k$. 
Corresponding phase diagram for the model suggests 
decrease in the tendency of abrupt fracture as number of components in the composite increase.
\end{abstract}


\maketitle

Fracture and material stability attracted human being from long time back when they 
started using materials found in nature to serve their own purposes. Fracture in materials 
is a complex phenomenon which involves very large length and time scales. The first scientific 
approach to realize the strength of materials was by Leonardo da Vinci 
\cite{Lund}, followed by 
Galileo Galilei and Edme Mariotte. During the last 25 years, fracture problems has been 
revisited by physicists \cite{Chak1, Herrmann, Kawa}. 
Fracture in materials are mainly guided by either extreme events or 
collective behavior of the cracks present in the material. In the later case, 
the material shows deformation in form of precursory rupture events prior to global failure. However, in the first case, 
a sharp crack, mostly the weakest one, spans the total length of the material 
denoting global failure without showing any precursory events. This behavior is disadvantageous 
for industrial purpose. 
This problem can be resolved to some extent 
by the introduction of composite materials
where materials of  
different mechanical and chemical properties are mixed to form a new material.
The composite material usually comes with either greater strength or with less 
risk of abrupt failure 
compared to its components \cite{brian,Luigi}. Example of such composites are cements, 
concrete, metal composites, ceramic 
composites etc. 

Fiber Bundle Model (FBM) is an important but simple model to study fracture-failure phenomena. 
This model is able to mimic correctly the avalanche behavior of fracture in materials \cite{Chak2}. 
It was first introduced by Pierce \cite{Pierce}. The FBM 
consists of fibers or Hookean springs, attached between two parallel plates. The plates are pulled apart by a force which 
exerts stresses on fibers. Once the stress crosses the breaking threshold of a particular fiber, 
chosen from a random distribution, the fiber breaks.  
There are two varieties of the FBM, equal load sharing FBM (ELS) and 
local load sharing FBM (LLS). For the ELS \cite{Pierce, Daniels}
no local deformation and hence no stress concentration occur anywhere around the failed fibers and 
the stress of the failed fibers is shared among all surviving fibers democratically. The load per
fiber therefore increases initiating failure of more fibers and so on. On the other hand, in LLS 
\cite{Phoenix1, Phoenix2, Phoenix3, Harlow1, Harlow2, Harlow3}, the terminal stress of a failed fiber 
is equally shared only among its 
nearest surviving neighbors.
  
In this paper, we have studied the fracture process of a 
composite bundle containing fibers of different elasticities with certain probabilities in the framework 
of fiber bundle model under ELS scheme. 
In \cite{kun2, kun3}, the authors considered a fiber bundle model where a fraction $\alpha$ are 
unbreakable and the remaining fibers are weak and have randomly distributed failure 
strength. The brittle to ductile transition was shown at a critical value 
of this fraction $\alpha_c$. 
In our model, there is no such unbreakable fraction of fibers. Here, 
inhomogeneity is introduced in terms of elastic modulus of the fibers. 
In case of a bundle having $k$-components, the elasticities of 
individual components are chosen between two extreme values $E_{min}$ and $E_{max}$ 
with spacing $\Delta E = (E_{max}-E_{min})/(k-1)$ and probability $1/k$. 
In a previous work, \cite{Kun} a system with continuously varying Young's modulus and with a constant 
breaking strength was studied.
However, in our model, elasticity 
is not the only source of disorder but also the 
threshold strength ($\sigma_i^{th}$) of fibers is. Here, the fibers have different thresholds chosen within the window 
$a$ to $a+2\delta$ of the uniform distribution $[0,1]$ where $a$ is the threshold 
stress corresponding to weakest link of the chain of $N$ fibers and 
$\delta$ is the half width of the distribution. Each fiber has a 
local elastic modulus $E(i)$. 
A strain $\epsilon$, same everywhere in the bundle, is applied 
externally. The load on the bundle at this condition is $\sigma=\langle E\rangle\epsilon$, where 
$\langle E\rangle = \frac{1}{N}\sum_{i=1}^N E(i)$. The breaking condition is
$\epsilon>\sigma_i^{th}/E(i)$ which depends both on elasticities and breaking thresholds. 
In this way we can say that fibers break when the externally applied strain crosses 
its local strain threshold $\epsilon_i^{th}=\sigma_i^{th}/E(i)$. After breaking of those fibers the strain is 
to be redistributed over the remaining fibers in equal amount 
keeping total stress constant. This may induce breaking of more fibers 
as the strain on the bundle gets increased.
At low $\delta$ values, 
the rupturing of the weakest fiber may result in failure of the entire system. However, at large $\delta$ values,
for the initial applied strain, 
the system will reach a stable point where no further rupture of fibers is possible as the next minimum strain threshold 
is above the redistributed strain.   
For further rupture, we have to 
increase the external strain by a little amount, 
$\epsilon \rightarrow \epsilon+\Delta\epsilon$ so that the breaking condition is satisfied. Thus the system will go through 
a series of stable states in succession of avalanches of rupturing of fibers till the critical strain
is reached
that can cause failure of the remaining fibers.

For analytical calculations, first we have chosen the 2-component system. 
If now the system suffers from a strain $\epsilon$,
the average stress of the system will be: 
$\sigma=\frac{1}{2}\epsilon(E_{1}+E_{2})=\frac{1}{2}(\sigma_1+\sigma_2)$, where $E_1=E_{min}$, $E_2=E_{max}$ are 
elasticities of fibers of type 1 and type 2 respectively and $\sigma_1, \sigma_2$ are stresses 
on type 1 and type 2 fibers respectively. 
At a certain point of this dynamics if $n_{b1}$ fraction of type 1 and $n_{b2}$ fraction of type 2 fibers are broken,
then the redistributed strain on the remaining fibers is
\begin{align}\label{eq:equation_n0}
\epsilon^{\prime}&=\epsilon + \frac{\displaystyle\sum_{j=1}^{2} n_{bj}\epsilon}{\left[1-\displaystyle\sum_{j=1}^{2} n_{bj}\right]}
\end{align}
Thus all fibers (type 1 and type 2) having breaking 
thresholds below $\epsilon^{\prime}$ will fail.

The corresponding stress 
on type $i$, i.e., $\sigma_{ri}$ is given by,
\begin{align}\label{eq:equation_n1}
\sigma_{ri}&=\sigma_i+ \frac{\displaystyle\sum_{j=1}^{2} n_{bj}\epsilon E_i}{\left[1-\displaystyle\sum_{j=1}^{2} n_{bj}\right]}
\end{align} 
where $i=1$ or $2$.

\noindent Broken fraction of type $i$ fibers will be
\begin{align}\label{eq:equation_nep}
n_{bi}=\frac{1}{2}\displaystyle\int^{\epsilon^{\prime}}_{a/E_i}p(\epsilon)d\epsilon
\end{align}
where $i=1$ or $2$ corresponding 
to the type of fibers and $p(\epsilon)$ is the strain threshold distribution. 
In terms of threshold stress distribution we have,
\begin{align}\label{eq:equation_n2}
n_{bi}=\frac{1}{2}\displaystyle\int^{\sigma_{ri}}_{a}p(\sigma)d\sigma
\end{align}
where $p(\sigma)$ is the uniform threshold distribution. 
Using Eq. \ref{eq:equation_n1} 
into Eq. \ref{eq:equation_n2} we get two quadratic equations for $n_{b1}$ and $n_{b2}$. 
Solving the quadratic equations we get the solutions for fraction of broken bonds as : 
\begin{align}
&n_{bi}=\displaystyle\frac{1}{2}\Bigg[\left(1-n_{bj}-\displaystyle\frac{a}{4\delta}\right) \pm \Bigg(\left(1-n_{bj}-\displaystyle\frac{a}{4\delta}\right)^2-\nonumber \\ &\displaystyle\frac{1}{\delta}\bigg(\sigma_{i}-a+an_{bj}\bigg)\Bigg)^{1/2}\Bigg]
\end{align}
where $i,j=1,2$ and $i \neq j$.

\noindent At critical point the solution becomes,  
\begin{align}
&n_{bi}^c=\displaystyle\frac{1}{2}\left(1-n_{bj}-\displaystyle\frac{a}{4\delta}\right)
\end{align}
Then total fraction broken at critical point for a particular width of disorder ($2\delta$) will be given as
\begin{align}
&n_b^c=\displaystyle\frac{1}{2}\left[2-\sum_{i=1}^2 n_{bi}^c-\displaystyle\frac{a}{2\delta}\right]  \nonumber \\
&\Rightarrow \frac{3}{2}n_b^c=\frac{1}{2}\left(2-\displaystyle\frac{a}{2\delta}\right)
\end{align}
Then fraction unbroken at critical point will be
\begin{align}\label{eq:equation1}
n_u^c=1-n_b^c=\frac{1}{3}n_b^c=\frac{1}{3}\left(1+\displaystyle\frac{a}{2\delta}\right)
\end{align}
At high $\delta$ value the system needs a continuous increment of strain to achieve global failure. 
Fraction unbroken before sudden failure in those cases are small. As we keep decreasing $\delta$ value $n_u^c$ 
increases and reaches $1$ at $\delta=\delta_c$, indicating an abrupt failure. 
Inserting $n_u^c=1$ we get, $\delta_c=\displaystyle\frac{a}{4}$.
For the uniform threshold distribution having its 
mean at $0.5$, $a=0.5-\delta$. We thus get the value of critical width: 
$\delta_c=0.1$. The existence of such critical disorder was discussed earlier in \cite{Sornette,Hansen}. 
For single component bundle $\delta_c$ was found to be $1/6$ in \cite{Subhadeep}.
With introduction of two types of fibers the critical point gets shifted to a lower value. We will 
discuss this point further while dealing with numerical results. 

To understand whether $\delta_c$ has any systematics with 
number of components used in the composite, we carried out a general 
treatment for a $k$ component bundle. Let at any instance $n_{b1}$ 
fraction of type 1, $n_{b2}$ fraction of type 2, $\cdots$, $n_{bk}$ fraction of type $k$ is broken. 
The corresponding stress on any fiber of type $j$ will be
\begin{align}
\sigma_{rj}=\sigma_j+\left[1-\displaystyle\sum_{i=1}^{k}n_{bi}\right]^{-1} \displaystyle\sum_{i=1}^{k}n_{bi}\sigma_j
\end{align}
where $\sigma_j$ is stress on type $j$th type. 
Proceeding in the same way as in the case of 2 components, we get 
the solution at critical point $n_{bj}^c$ as,
\begin{align}
n_{bj}^c=\frac{1}{2}\left[1-\displaystyle\sum_{i \ne j;i=1}^{k}n_{bi}-\displaystyle\frac{a}{2k\delta}\right]
\end{align}
which finally gives the fraction unbroken for a k-component system as
\begin{align}\label{eq:equation2}
n_u^c=1-n_b^c=\displaystyle\frac{1}{k+1}\left(1+\displaystyle\frac{a}{2\delta}\right)
\end{align}
In above equation taking $n_u^c=1$, we get $\delta_c$ in terms of $a$ below which the 
model shows abrupt failure.
\begin{align}\label{eq:equation3}
\delta_c=\displaystyle\frac{a}{2k}.
\end{align}
For the uniform threshold distribution with mean at $0.5$, we have from Eq. \ref{eq:equation3}
\begin{align}\label{eq:equation4}
\delta_c=\displaystyle\frac{0.5}{2k+1}.
\end{align}
This result clearly indicates that the tendency of abrupt failure decreases with increasing number of components.

\begin{figure}[ht]
\centering
\captionsetup{justification=raggedright}
\includegraphics[width=6cm]{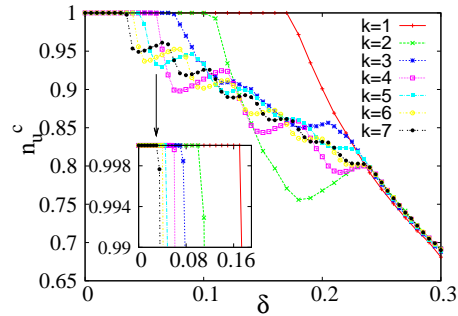}
\caption{(Color online) Variation of $n_u^c$ with $\delta$ for $E_{max}-E_{min}=1.0$ and different values of $k$. 
The region of $\delta$ where $n_u^c$ starts to deviate from $1$ by an appreciable amount is being zoomed and shown 
in the inset. The zoomed part clearly shows the decrease in $\delta_c$ with increasing $k$.}
\label{fig:Critical_Point_ManyElasticity}
\end{figure}

To have some better understanding for the transition, we have also studied the problem numerically. 
For numerical simulations we have considered a bundle of $L$ 
fibers with their strengths chosen from a uniform distribution. 
\begin{figure}[ht]
\centering
\captionsetup{justification=raggedright}
\includegraphics[width=8.5cm]{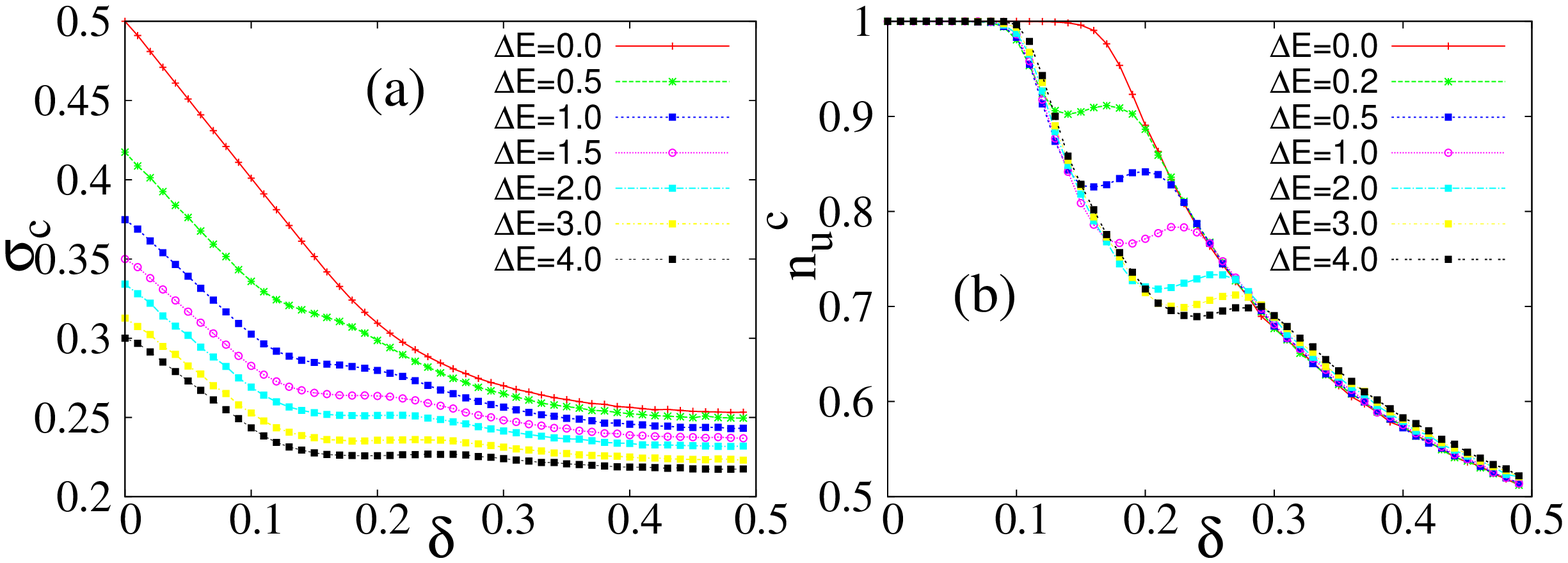}
\caption{(Color online) Study of (a) $\sigma_c$ and (b) $n_u^c$ for different $\Delta E$ keeping $E_{min}=1.0$ for $k=2$. 
Here $L=5\times10^4$. $\Delta E=0$ is the case for conventional fiber bundle model 
with only one type of fiber. }
\label{fig:Critical_Point}
\end{figure}
The critical average stress $\sigma_c$ is defined as the average stress on the fibers corresponding 
to a particular $\delta$ 
that can cause failure of all surviving fibers.
For only one type of material present ($\Delta E=0$), $\sigma_c$ starts from 
$1/2$ at $\delta=0$, decreases with increasing $\delta$ values and 
reaches $1/4$ at  $\delta=1/2$ \cite{Chak2}. 
Now if we mix two types of fibers with elasticities $E_1=E_{min}$ (chosen $1.0$ throughout) and 
$E_2=E_{max}$, then the profile of 
$\sigma_c$ gets lowered at any 
$\delta$ value for $\Delta E>0$ compared to the case of $\Delta E=0$. 
This suggests that the strength of the 
material decreases as 
we mix two types of fibers, which apparently seems to be a disadvantage from the perspective of engineering 
science. However, the actual advantage of such mixing lies within the behavior of 
$n^c_u$ which is already claimed in the analytical calculations.  
From numerical data, as shown in Fig. \ref{fig:Critical_Point} we can easily observe 
that as we mix two types of fibers, $\delta_c$ has a value close to $0.1$ instead of $1/6$ (for one type of fiber only). 
Thus brittle region for the two-component composite
is smaller as $n_u^c$ starts decreasing from $1$ beyond $\delta=0.1$. Thus unlike a one-component 
material, here for a greater range of $\delta$, the system behaves as a quasi-brittle material 
which is favorable for practical use. 
Beyond a certain $\delta$, there is a tug-of-war  between brittle and quasi brittle behavior,
for which $n_u^c$ increases very slowly for a range of $\delta$. 
This is perhaps due to the fact that for a particular value of strain acting 
on a two-component system, the fiber with greater elasticity 
experiences a local stress greater than the average stress $\sigma$ and therefore 
has a greater chance for failure.
For more than two components present in the composite, variation of 
$n_u^c$ with $\delta$ is shown in Fig. \ref{fig:Critical_Point_ManyElasticity}. 
For very high values 
of $\delta$, $n_u^c$ remains unaffected by number of components at a particular $\Delta E$ value. For low 
$\delta$, however, there is a huge change. As we increase the varieties of fibers,
$\delta_c$ value get shifted to lower and lower values. The region close to $n_u^c=1$ is further 
zoomed in the inset and it clearly  
shows the decrease in $\delta_c$ with increase in $k$. The numerical results agree well with analytical results. 
Reduction of $\delta_c$ suggests 
decrease in the fracture abruptness and makes the model more effective for practical use. Beyond $\delta_c$, 
for $k > 1$, there are plateau type regions, number of which is $k$. This indicates again a competition 
between abrupt and continuous fracture due to different stress levels 
for different types of fibers. A situation $n_u^c=1$ corresponds 
to an abrupt failure where for the applied strain all the fibers break
in a single step. This holds for a $\delta \leq \delta_c$.
When we increase the strain from zero, up to the point of no fiber breaking, the strain 
($\epsilon$) shows a linear relationship with average stress ($\sigma$). With increasing strain when 
one fiber breaks the total model breaks down and the 
linear relationship of strain and stress per fiber holds for the whole time. 
\begin{figure}[ht]
\centering
\captionsetup{justification=raggedright}
\includegraphics[width=6cm]{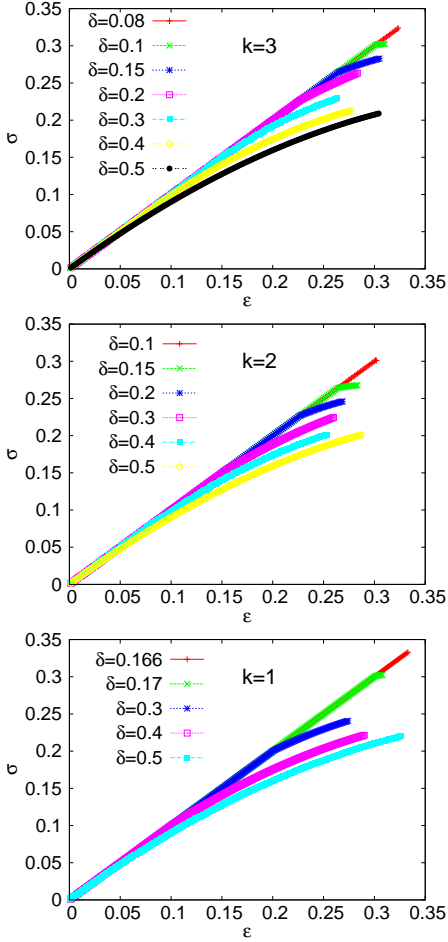}
\caption{(Color online) Study of stress-strain relationship for the bundle of fibers for $k=1$ (below), $k=2$ 
(middle) and $k=3$ (top). System size, $L=10^4$. Beyond $\delta_c$ the response curves become nonlinear.
} 
\label{fig:ResponseCurve}
\end{figure}
Now when $n_{u}^{c} < 1$, i.e., $\delta > \delta_c$
there are a number of stable points before global failure and between any two consecutive stable points there are a 
number of redistributing steps. 
Due to such redistribution the strain before and after any 
stable point deviates from each other. Because of this deviation the linear behavior in between applied 
stress (which is a function of applied strain) and strain in the model (redistributed strain) is lost. The response 
curve in therefore non-linear. This is shown in Fig. \ref{fig:ResponseCurve}.

\begin{figure}[ht]
\centering
\captionsetup{justification=raggedright}
\includegraphics[width=8cm]{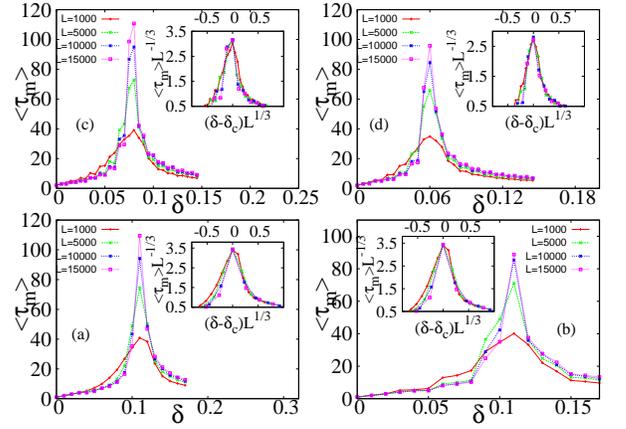}
\caption{(Color online) Study of $\langle\tau_m\rangle$ against $\delta$ for 
(a) $k=2, \Delta E=1.0$; (b) $k=2, \Delta E=2.0$; (c) $k=3, \Delta E=1.0$  
and (d) $k=4, \Delta E=1.0$ with $L=1000$(plus), $5000$(cross), $10000$(star), $15000$(square). 
In the inset the system size scaling for $\langle\tau_m\rangle$ is shown:
$\langle\tau_m\rangle \sim L^{\gamma/\nu}\Phi\big((\delta-\delta_c)L^{1/\nu}\big)$
with $\gamma=1$ and $\nu\approx3$ 
for different values of $\delta_{c}$ depending on number of components.} 
\label{fig:Relaxation_Time}
\end{figure}

The exact value of $\delta_c$ can be confirmed from finite size scaling of relaxation time as is done in \cite{new1, new2}. 
Relaxation time ($\tau$) in the model is defined as the number of redistributing 
steps after the application of the minimum stress corresponding to the weakest link. 
However, due to system size effect there will be some configurations 
where even for $\delta < \delta_c$, the model would not reach failure point. To avoid this
we have calculated the maximum possible relaxation time ($\tau_m$) from a set of $10^3$ configurations. 
Averaging over such $10^3$ 
$\tau_m$ values we get the average maximum relaxation time 
$\langle\tau_m\rangle$ which is plotted against $\delta$ for different $L$ in Fig. \ref{fig:Relaxation_Time}. 
$\langle\tau_m\rangle$ shows a peak at $\delta_c(L)$.
This  
approach was used earlier in \cite{Subhadeep}.
To understand the system size effect, finite size scaling is done where  
\begin{align}\label{eq:equation_scale}
\langle\tau_m\rangle \sim L^{\gamma/\nu}\Phi\big((\delta-\delta_c)L^{1/\nu}\big).
\end{align}
After the scaling, all the plots collapse for exponent values, $\gamma=1$ and $\nu\approx3$. 
Close to $\delta_c$, 
\begin{align}
\langle\tau_m\rangle \sim (\delta-\delta_c)^{-\gamma}
\end{align}
Following this approach for a two-component system, we have found that 
the peak gets shifted close to $0.11$ from 
$1/6$ This is shown in Fig. \ref{fig:Relaxation_Time} for two different $\Delta E$ values $1.0$ and $2.0$, 
(Fig. \ref{fig:Relaxation_Time}(a) and Fig. \ref{fig:Relaxation_Time}(b)). 
It is clear from Fig. \ref{fig:Relaxation_Time} (a and b) that, the critical behavior 
is not affected by the difference in elasticities of the two components but solely 
on the number of components in the composite.        
$\delta_c$ for $k>2$ is further analyzed from the study of $\langle\tau_m\rangle$. 
$\langle\tau_m\rangle$ is calculated for $\Delta E=1.0$ and for $L=1000$, 
$5000$, $10000$ and $15000$. The results for $k=3,4$ are shown in Fig. \ref{fig:Relaxation_Time}(c) 
and Fig. \ref{fig:Relaxation_Time}(d).  
As $k$ value changes $\delta_c$ is tuned properly following the same scaling rule (Eq. \ref{eq:equation_n2}) 
to obtain collapse. These give the proper $\delta_c$ values for a particular set. For higher $k$, again value 
of $\delta_c$ is lower.

\begin{figure}[ht]
\centering
\captionsetup{justification=raggedright}
\includegraphics[width=6cm]{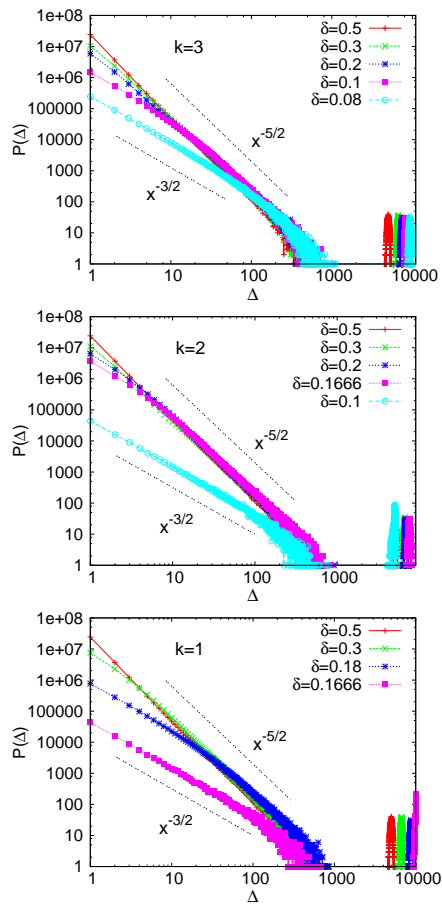}
\caption{(Color online) The burst size distribution $P(\Delta)$ of a bundel with $10^4$ fibers for $k=1$ (bottom), $2$ (middle), $3$ (top). 
For $\delta \leq \delta_c$, $\alpha$ is found to be 
$3/2$ but when 
$\delta > \delta_c$, $\alpha$ shows a value $5/2$.
} 
\label{fig:BurstSize}
\end{figure}

The $\delta_c$ values can also be confirmed from the study of burst size distributions. 
If the external strain is raised in such a way that it
becomes equal to the breaking threshold of the weakest
fiber, then the fiber breaks. This triggers an avalanche of
fiber failures and the bundle may attain
a new stable state. The total number $\Delta$ of fibers that
fail in this event is called the burst size. Starting
from a set of intact fibers, the global failure
can be attained by raising the external strain. The burst size distribution $P(\Delta)$ is shown in Fig. 
\ref{fig:BurstSize} for $k=1$ (bottom), $2$ (middle), $3$ (top). The distribution shows a 
power law as $P(\Delta) \sim \Delta^{-\alpha}$, where $\alpha$ is the exponent. This is in agreement with \cite{Hemmer}
where $\alpha$ was shown to be $5/2$. In \cite{manna}, the authors showed that for
power law distributed breaking thresholds $\alpha$ shows a crossover at $\Delta_c$, where below $\Delta_c$
$\alpha$ is $3/2$ and above $\Delta_c$, $\alpha$ is $5/2$. In our case we have a uniform threshold distribution, however,
the window of distribution $\delta$ is varying around $0.5$.
For $\delta=\delta_c$, $\alpha$ is found to be 
$3/2$ but when $\delta > \delta_c$, $\alpha$ shifts to a value $5/2$. 
This sudden change in $\alpha$ around critical disorder confirms $\delta_c=0.1666, 0.1$ and $0.08$ 
respectively for $k=1, 2$ and $3$.

Now for such a model the window $0>\delta>\delta_c$ 
always possesses risk of abrupt failure. Therefore, the model can be used safely only with disorder width $\delta>\delta_c$ 
within the interval [0,1]. Thus we can define a effective working region (EWR) 
for the material, given by: EWR = $(1.0-2\delta_c)\times 100\%$. 
Composite materials show a huge reduction in the fracture abruptness.
As a result, EWR is also increased as shown in the following Table \ref{table : kysymys}. 

\begin{table}[h]
\begin{tabular}{|c|c|c|c|c|}
 \hline
 $k$ & \multicolumn{2}{ c| }{$\delta_c$} & \multicolumn{2}{ c| }{$EWR (\%)$} \\ \cline{2-5}
 & Analytical & Numerical & Analytical & Numerical \\ \hline
 $1$ & $1/6$ & $1/6$ & $66.8$ & $66.8$ \\ \hline
 $2$ & $0.10$ & $0.108$ & $80.0$ & $78.4$ \\ \hline
 $3$ & $0.071$ & $0.077$ & $85.8$ & $84.6$ \\ \hline
 $4$ & $0.055$ & $0.060$ & $89.0$ & $88.0$ \\ \hline
 $5$ & $0.045$ & $0.049$ & $91.0$ & $90.2$ \\ \hline
 $6$ & $0.038$ & $0.041$ & $92.4$ & $91.8$ \\ \hline
\end{tabular}
\captionsetup{justification=raggedright}
\caption{Analytical and numerical values of $\delta_c$ and working regions for different $k$.}
\label{table : kysymys}
\end{table}

\begin{figure}[ht]
\centering
\captionsetup{justification=raggedright}
\subfloat{\includegraphics[width=6cm, keepaspectratio]{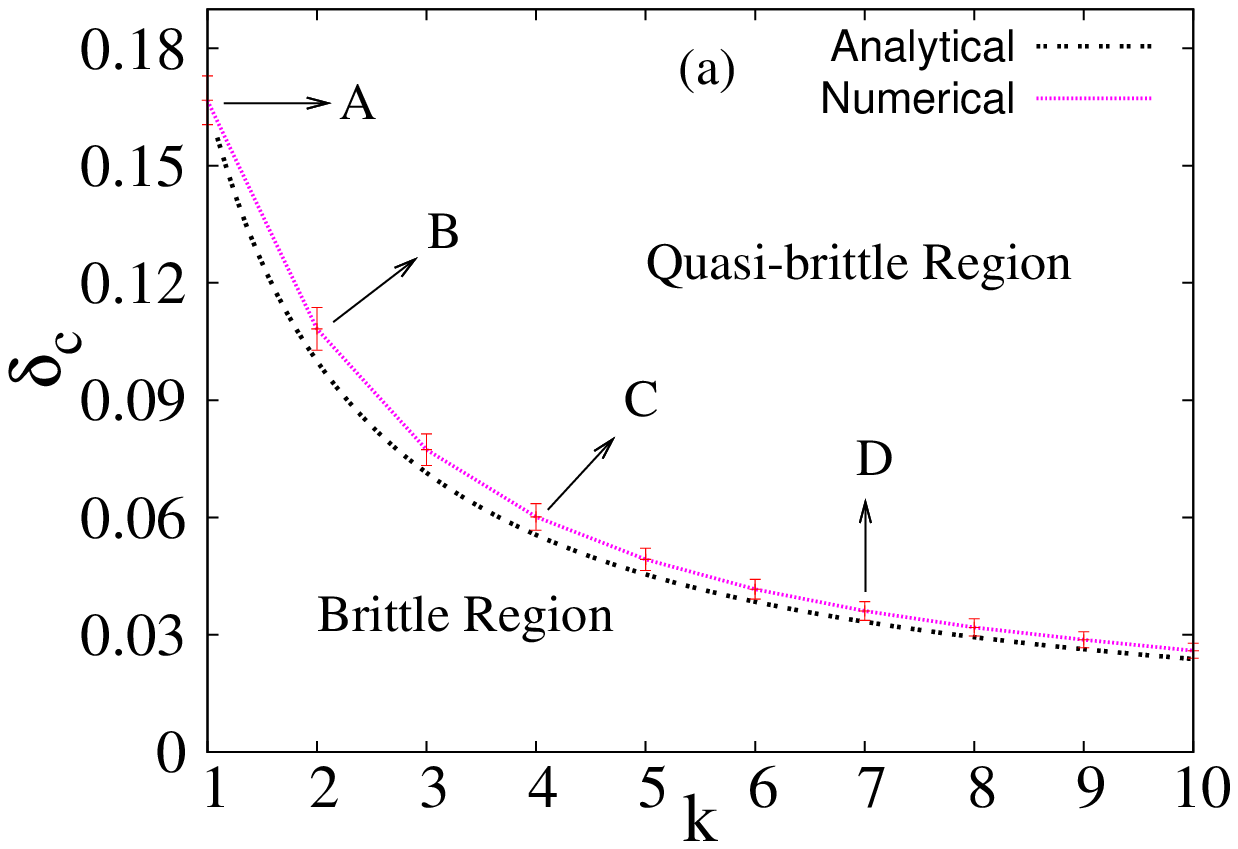}} \\ 
\subfloat{\includegraphics[width=6cm, keepaspectratio]{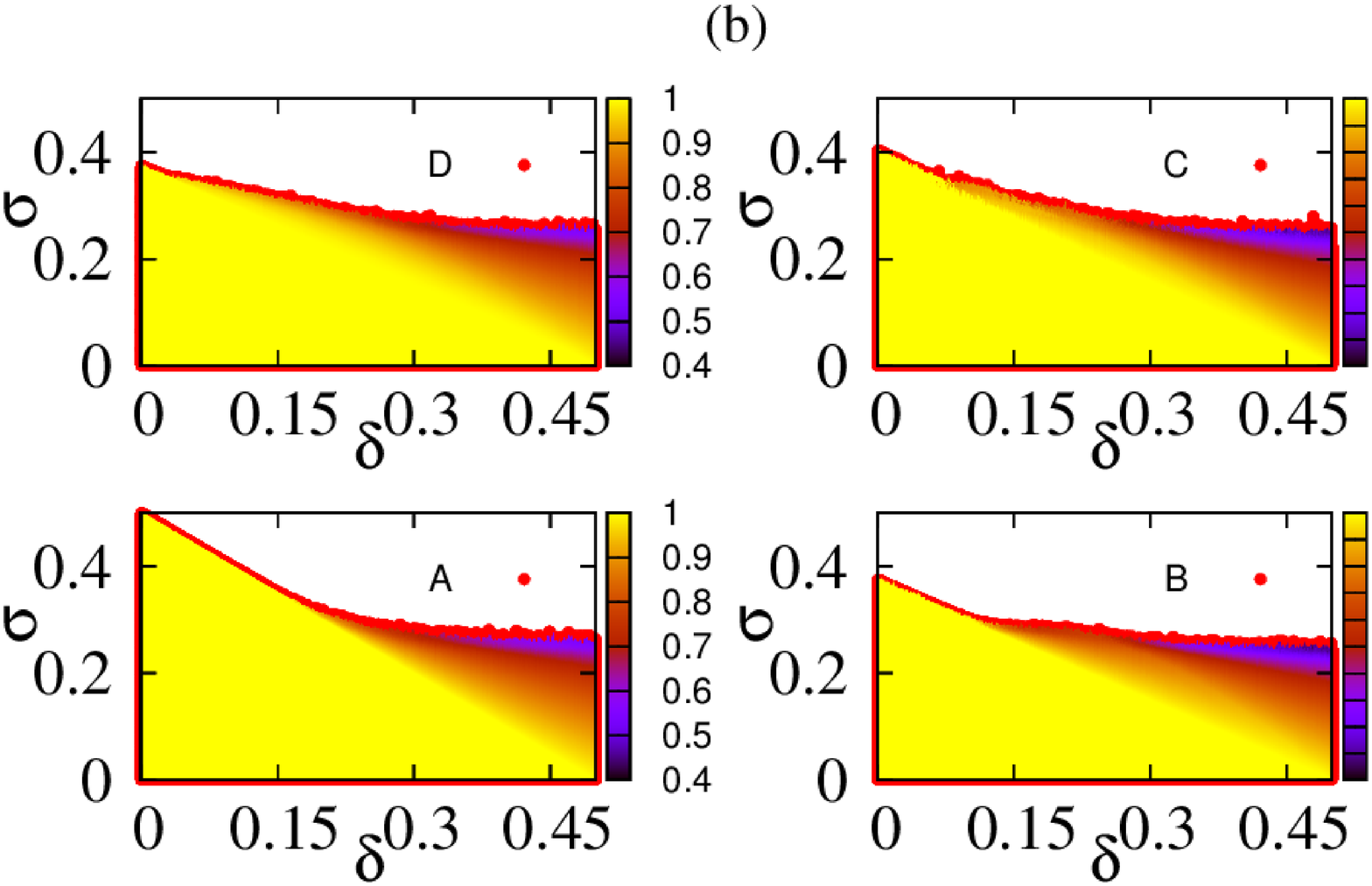}}
\caption{(Color online) (a) Phase diagram of the model for both 
numerical (pink) and analytical (black dashed dotted) results. 
The points $A$, $B$, $C$ and $D$ corresponds to $k$ values 
$1, 2, 4$ and $7$. (b) Surface plot for A, B, C, D 
where fraction of unbroken fibers $n_u$ are plotted for different $\sigma$ and $\delta$.
For $\delta < \delta_c$, fracture is abrupt, one can cross yellow region by a single jump. 
For $\delta < \delta_c$, fracture is gradual, color gradient should be crossed after the yellow region.}
\label{fig:Phase_Diagram}
\end{figure}
Next we plot $\delta_c$ against $k$, the phase diagram of our model (Fig. \ref{fig:Phase_Diagram}(a))
in form of a transition line below which 
the system is brittle, not suitable for practical purpose and above quasi-brittle, suitable for use. 
If we increase $k$ value, the brittle region gradually 
decreases and quasi-brittle region increases which in turn increases the EWR. 
In Fig. \ref{fig:Phase_Diagram}(b) 
we give the surface plot for points $A$, $B$, $C$ and $D$ in the phase diagram 
corresponding to $k$ values $1, 2, 4$ and $7$. For different $\delta$ values within the range $0$ to $0.5$ (along X-axis) we have 
calculated fraction of unbroken fibers $n_u$ for a set of average stress $\sigma$ starting from the minimum stress 
corresponding to the weakest link to the critical one (along Y-axis).  
Yellow region corresponds to initial configuration where all fibers are intact while the color gradient represents 
the partially broken phase. For $\delta<\delta_c$ the model breaks abruptly and we can simply 
jump out of the yellow region crossing a straight line. For $\delta>\delta_c$ the model 
undergoes a continuous fracture so that one has to cross the color gradient after the yellow region.  
The surface plot clearly supports the decrease in $\delta_c$ with increasing number of components ($k$) 
shown in the phase diagram. 

In conclusion, 
our work gives a good design aspect for materials for practical use. 
Under ELS scheme of fiber bundle model, we see a critical behavior at a 
critical value of threshold distribution 
width ($\delta_c$) which decreases as we increase the number of components in the composite. 
This $\delta_c$ serves as a 
transition point for a material to behave like brittle or quasi-brittle. 
If we mix a number of components, the composite shows decreased fracture abruptness 
with some systematics. This in turn helps increasing the effective working region for the material.
This model, in this way, may help designing good composite materials 
although good composites should have other properties like 
light weight, high stiffness etc., studies of which are beyond the scope of this paper.

Acknowledgement : The authors thank Purusattam Ray for some delightful comments and discussions, and 
Soumyajyoti Biswas for critical reading of the manuscript.




\begin{thebibliography}{99}
\bibitem{Lund} J. R. Lund and J. P. Byrne, Civ. and Eng. Environ. Syst. {\bf 18}, 243 (2001).
\bibitem{Chak1} B. K. Chakrabarti and L. G. Benguigui, {\it Statistical Physics of Fracture and 
Breakdown in Disordered Systems} (Oxford University Press, Oxford, 1997).
\bibitem{Herrmann} {\it Statistical Models for the Fracture of Disordered Media} edited by H. J. Herrmann and S. Roux 
(North Holland, Amsterdam, 1990).
\bibitem{Kawa} H. Kawamura, T. Hatano, N. Kato, S. Biswas and B. K. Chakrabarti, Rev. Mod. Phys. {\bf 84}, 839 (2012).
\bibitem{brian} B. Lawn, {\it Fracture of brittle Solids} (Cambridge University Press, Cambridge, 1993).
\bibitem{Luigi} L. Nicolais, A. Borzacchiello and S. M. Lee, {\it Wiley Encyclopedia of Composites} (Wiley online library, 2012).
\bibitem{Chak2} S. Pradhan, A. Hansen and B. K. Chakrabarti, Rev. Mod. Phys. {\bf 82}, 499 (2010).
\bibitem{Pierce} F. T. Pierce, J. Text. Ind. {\bf 17}, 355 (1926).
\bibitem{Daniels} H. E. Daniels, Proc. R. Soc. London, Ser. A {\bf 183}, 405 (1945).
\bibitem{Phoenix1} S. L. Phoenix, Adv. Appl. Probab. {\bf 11}, 153 (1979).
\bibitem{Phoenix2} R. L. Smith and S. L. Phoenix, J. Appl. Mech. {\bf 48}, 75 (1981). 
\bibitem{Phoenix3} W. I. Newman and S. L. Phoenix, Phys. Rev. E {\bf 63}, 021507 (2001).
\bibitem{Harlow1} D. G. Harlow and S. L. Phoenix, J. Compos. Mater. {\bf 12}, 314 (1978). 
\bibitem{Harlow2} D. G. Harlow and S. L. Phoenix, Adv. Appl. probab. {\bf 14}, 68 (1982). 
\bibitem{Harlow3} R. L. Smith, Proc. R. Soc. London, Ser. A {\bf 382}, 179 (1982).
\bibitem{kun2} R. C. Hidalgo, K. Kov\'{a}cs, I. Pagonabarraga and F. Kun, Europhys. Lett. {\bf 81}, 54005 (2008).
\bibitem{kun3} K. Kov\'{a}cs, R. C. Hidalgo, I. Pagonabarraga and F. Kun, Phys. Rev. E {\bf 87}, 042816 (2013).
\bibitem{Kun} E. Karpas and F. Kun, Europhys. Lett. {\bf 95}, 16004 (2011).
\bibitem{Sornette} J. V. Andersen,  D. Sornette and K. T. Leung, Phys. Rev. Lett. {\bf 78}, 2140 (1997).
\bibitem{Hansen} S. Pradhan and A. Hansen, Phys. Rev. E {\bf 72}, 026111 (2005).
\bibitem{Subhadeep} S. Roy and P. Ray, arXiv:{\bf 1412.1211}.
\bibitem{new1} S. Pradhan, P. Bhattacharyya, and B. K. Chakrabarti, Phys. Rev. E {\bf 66}, 016116 (2002). 
\bibitem{new2} C Roy, S Kundu, and S. S. Manna, Phys. Rev. E {\bf 87}, 062137 (2013).
\bibitem{Hemmer} P. C. Hemmer and A. Hansen, J. Appl. Mech. {\bf 59}, 909 (1992).
\bibitem{manna} C. Roy, S. Kundu and S. S. Manna, Phys. Rev. E {\bf 91}, 032103 (2015).
\end{thebibliography}
\end{document}